# Quantization of The Three Dimensional Sinai Billiard

Harel Primack and Uzy Smilansky

*Department of physics of complex systems,*

*The Weizmann Institute, Rehovot 76100, Israel*

(February 1, 1995)

## Abstract

For the first time a three–dimensional (3D) chaotic billiard – the 3D Sinai billiard – was quantized, and high–precision spectra with thousands of eigenvalues were calculated. We present here a semiclassical and statistical analysis of the spectra, and point out some of the features which are genuine consequences of the three dimensionality of this chaotic billiard.

03.65.Sq, 05.45.+b

Typeset using REVTEX

Most of the work in the field of "quantum chaos" is related to systems with two degrees of freedom. This is a natural tendency, since two–dimensional (2D) systems provide the simplest, non–trivial examples of chaotic systems. In this respect, 2D billiards were subjected to thorough studies and served as good paradigms for the understanding of time–independent chaotic Hamiltonian systems.

The main motivation to extend the study of quantum chaotic billiards to three dimensions, is to examine the semiclassical approximation: this approximation introduces an inherent error of order $\hbar^2$ to the energy levels (regardless of dimension) [1], while the mean energy spacing is proportional to $\hbar^d$, for systems with $d$ degrees of freedom [2]. This might prohibit the resolution of single (high–lying) levels when $d > 2$. In the absence of a complete theory, it is important to test these ideas, to check whether the semiclassical approximation can be recovered by suitable $\hbar$–expansions [3,4], and to evaluate its usefulness for statistical purposes. Other motivation to study 3D systems is the existence of "real world" systems that cannot be reduced into 2D idealizations, and which can exhibit some (classical and/or quantum) genuine 3D effects. Also, for $d > 2$, classical systems can support all the possible types of stability, including mixed (lexodromic) stability, which is absent in 2D systems.

In this Letter, we would like to report about the numerical quantization and spectral analysis of the 3D Sinai billiard. The results will be presented with special attention to genuine 3D features which affect the spectrum in whatever way it is approached and analyzed. To the best of our knowledge, this is the first study of 3D quantum chaotic billiards which can match the studies of 2D billiards in accuracy and extent. We study 3D Sinai billiard, which is the space inside a cube of edge $L$ and outside an embedded sphere of radius $R$, $R < L/2$ (see fig. 1). The classical motion consists of a particle that is specularly reflected from the sphere and from the the cube's walls.

The 3D Sinai billiard was quantized using the KKR method [5–7] which was previously used for the quantization of the 2D Sinai billiard [8]. The essence of the method is to transform the Lippmann–Schwinger integral equation into a particularly efficient system of linear equations. This is obtained by a special choice of the basis functions and exploiting the symmetries of the problem: the lattice periodicity and the spherical symmetry of the inscribed potential. This reduces the numerical effort considerably. The method heavily relies on a precise and efficient calculation of "lattice constants" [6] which is achieved by using the full Ewald summation technique [7,9], and choosing the convergence parameter in an optimal way which leads to the fastest convergence for a prescribed accuracy. Details of the calculation will be given elsewhere [10].

The Hamiltonian of the full 3D Sinai billiard is invariant under the discrete group $O_h$ of the cube's rotations and inversion [11]. This group consists of 48 elements and 10 classes, thus the quantum system suffers from geometrical degeneracy, and the spectrum is composed of 10 independent spectra. In order to remove this "trivial" degeneracy [8], we desymmetrized the billiard by using as basis functions linear combinations of spherical–harmonics that transform according to one of the irreducible representations (irreps) of $O_h$ ("cubic harmonics" [12]). We found a simple, non-recursive method for determining the cubic harmonics, using the singular value decomposition (SVD) algorithm [13], and calculated the coefficients for the completely symmetric ($A1$) and completely antisymmetric ($A1'$) one–dimensional irreps. These correspond to introducing Neumann and Dirichlet boundary conditions, respectively, on the symmetry planes of the cube, forming a desymmetrized element which is a triangular pyramid (it is marked by heavy lines in fig. 1). It is important to emphasize, that the above decomposition of the Hamiltonian is crucial for the feasibility of the calculation: had the $O_h$ symmetry not exist, we would have to deal with matrices larger by a factor of $48^2$.

We calculated few independent spectra of the billiard for $L = 1$ and $R = 0.2, R = 0.3$. Dirichlet and Neumann boundary conditions were applied independently on the sphere and on the symmetry planes. The spectrum for $R = 0.2$ and Dirichlet boundary conditions on both sphere and planes contained 2288 levels and covered the range $0 \leq k_n \leq 200$ ($k_n^2 \equiv E_n$). This is the largest spectral interval obtained so far, and we shall refer to it as $R = 0.2$ (DD). The other three spectra for the same radius contained 1000 levels each, and the spectra for $R = 0.3$ contained 760 levels each. The completeness of the data was checked by comparing the actual number of eigenvalues $N(k)$ to Weyl's law $\bar{N}(k)$ [14], in order to detect either missing or redundant eigenvalues [15]. For all the spectra computed, $N(k) - \bar{N}(k)$ fluctuated about 0 with no systematic deviations, indicating with high probability the completeness of the spectra. A further consistency check of the spectrum is obtained by comparing the lower eigenvalues to the eigenvalues of the $R = 0$ integrable (and separable) billiard. For Dirichlet boundary conditions on the planes, the lowest angular momentum that appears in the KKR determinant is $l = 9$. Thus, for $kR \lesssim 9$ we expect the eigenvalues of the billiard to be very close to the $R = 0$ eigenvalues, which is indeed confirmed by the numerical results. We performed also some convergence checks to estimate the error in the eigenvalues, and found it to be of order $10^{-4}$ of the mean level spacing.

In order to facilitate later analysis, it is instructive to examine first the various classes of the classical periodic manifolds that exist in the 3D Sinai billiard, since they constitute the building blocks of the semiclassical analysis. First, we have the generic, isolated and unstable periodic orbits (1–dimensional manifolds). They are typical to hyperbolic systems in any dimension, and their total number proliferates exponentially with their length [2]. For our billiard, they are characterized by at least one non–tangential collision with the sphere. Second, we have the non–generic, neutrally stable, "bouncing ball" orbits [16,8,17,18], which bounce only between symmetry planes and/or the walls of the cube. For the specific case of the desymmetrized 3D Sinai billiard, they are conveniently analyzed by considering the integrable $R = 0$ case. The quantum density of states can be explicitly written in this case as:

$$\begin{aligned} d_{R=0}(E) &= \sum_{n>l>m>0} \delta\left[E - \left(\frac{2\pi}{L}\right)^2 (n^2 + l^2 + m^2)\right] \\ &= \frac{L^3 \sqrt{E}}{192\pi^2} \sum_{pqr} \mathrm{sinc}\left(L\sqrt{p^2 + q^2 + r^2}\sqrt{E}\right) - \frac{L^2}{64\pi} \sum_{pq} J_0\left(L\sqrt{p^2 + q^2}\sqrt{E}\right) \\ &\quad - \frac{L^2}{32\sqrt{2}\pi} \sum_{pq} J_0\left(L\sqrt{p^2 + \frac{q^2}{2}}\sqrt{E}\right) + \frac{3L}{32\pi\sqrt{E}} \sum_p \cos\left(Lp\sqrt{E}\right) \\ &\quad + \frac{L}{16\sqrt{2}\pi\sqrt{E}} \sum_p \cos\left(\frac{L}{\sqrt{2}}p\sqrt{E}\right) + \frac{L}{12\sqrt{3}\pi\sqrt{E}} \sum_p \cos\left(\frac{L}{\sqrt{3}}p\sqrt{E}\right) - \frac{5}{16}\delta(E-0) \quad (1) \end{aligned}$$

where $\mathrm{sinc}(x) \equiv \sin(x)/x$, and $p, q, r \in \mathbb{Z}$. The oscillatory part of the first term

$((p, q, r) \neq (0, 0, 0))$ describes contributions from periodic manifolds (tori) that occupy 3D volumes in configuration space. Each such manifold is a (2–parameter) continuous family of periodic orbits of the same length. The other oscillatory contributions come from 1D and 2D manifolds of periodic orbits that are confined to special lines and planes, respectively. They involve collisions with corners and edges, and need special care which will be given elsewhere [10]. It is interesting to note, that isolated and neutral periodic orbits (1D manifolds) exist, and in fig. 1 we show two of them, that contribute to the fifth and sixth terms of (1). The triangular face $ABC$ supports an infinite set of 2D manifolds that stay on it, as if it were a 2D triangle. The number of bouncing ball manifolds grows as a power of their length, in sharp contrast to the exponential growth of the generic unstable orbits. For $R > 0$, infinitely many periodic manifolds are either completely or partially pruned. The remaining periodic manifolds are "terminated" with periodic orbits that are tangent to the sphere. These tangential manifolds constitute the third kind of periodic manifolds, and they are of one dimension lower than the bouncing ball manifold that they terminate. A comparison with the situation in 2D Sinai billiard reveals a substantial difference: there is only a finite number of bouncing ball manifolds in the 2D case $(0 < R < L/2)$ [8], in contrast to the infinite number in the 3D case. Also, only in 3D case there exist 3D bouncing ball manifolds and 2D tangential manifolds. These differences emphasize the almost unavoidable rôle of the bouncing balls in the 3D case, which is a direct consequence of the high symmetry of the original, non–desymmetrized billiard. Thus, the same high symmetry that allowed an efficient decomposition of the Hamiltonian, is responsible for the prominent non–generic bouncing ball manifolds.

In the sequel, we shall study the relations between the quantum spectra and the classical periodic manifolds, in light of the discussion above. The first step is to construct the length spectrum, which is the (finite) cosine–transform of the oscillatory (wavenumber) density of states:

$$D(x; k) \equiv \frac{1}{k} \int_0^k \mathrm{d}k' \cos(k'x) d_{osc}(k'). \tag{2}$$

Semiclassically, $D(x;k)$ is expected to be peaked near lengths $x$ that correspond to classical periodic manifolds of the billiard [16,17], thus, allowing us to extract classical information out of the quantum data. Substituting (1) into (2), we get that the contribution of a $d$-dimensional manifold to $D(x;k)$ is $O(k^{(d-1)/2})$. This is also valid for the 1D unstable periodic orbits [16]. Hence, the (leading, 3D) bouncing ball contributions are expected to be much stronger ($O(k)$) than those of the isolated periodic orbits. This is to be contrasted with the $O(k^{1/2})$ enhancement of bouncing balls in the 2D case [16]. The length spectrum of the $R = 0.2$ (DD) spectrum is plotted in fig. 2 for $k = 100$ and $k = 200$, and the lengths of the bouncing ball orbits for $0 \leq x \leq 2$ are marked together with their expected $k$-dependence. We clearly observe peaks for the predicted lengths (with possible exception of 1D contribution at $x = 2/\sqrt{3} \approx 1.15$ due to interferences), and qualitative agreement with the predicted $k$-dependence. The bouncing balls are clearly seen to dominate $D(x;k)$. Our only hope to detect an unstable periodic orbit is to look in the domain where the contributions are still fairly isolated, $x \lesssim 0.6$. The shortest unstable periodic orbit, along the edge $AE$ of fig. 1, has length of 0.6 for the $R = 0.2$ case, which is very close to the length $1/\sqrt{3} \approx 0.577$ of the shortest 1D bouncing ball. For $R = 0.3$ the situation is better, because the shortest unstable periodic orbit has length of 0.4, which is well separated from the shortest 1D bouncing ball. In both cases, however, the expected contribution is not observed. A detailed calculation of the semiclassical contribution of this unstable periodic orbit reveals, that being on the axis of an 8-fold symmetry greatly diminishes the semiclassical contribution of this orbit with respect to the situation where desymmetrization is not imposed. Quantitatively, the ratio is given by:

$$\rho_8(\beta) = \frac{1}{4}\left[2 - 2\sqrt{1-2\beta} - \beta\left(\frac{2-\beta}{1-\beta}\right)\right], \quad \beta \equiv \frac{R}{L} \qquad (3)$$

$$\approx \left(\frac{\beta}{2}\right)^4 \quad \text{for } \beta \to 0$$

In particular, $\rho_8(R/L = 0.2) \approx 2 \times 10^{-4} \ll 1$. We emphasize that this is a purely 3D geometrical effect due to desymmetrization, and that the amplitude of this contribution to $D(x;k)$ is $k$–independent as for all other unstable periodic orbits.

The above analysis shows, that if one is interested in the study of the generic contributions, it is imperative to develop a method which eliminates the non–generic contributions in an accurate and simple way. One possible idea is to parameterize the boundary conditions on the sphere, and use this freedom as a selection filter to extract the contributions of generic periodic orbits exclusively. Work along these lines is now in progress [19].

We applied some of the common spectral statistics to the spectra described above. In fig. 3 we present the nearest neighbor distribution $P(s)$ for the $R = 0.2$ and $R = 0.3$ spectra. Both curves are similar to the GOE curve, but for $R = 0.2$ there are noticeable deviations for small $s$. These deviations are attributed to the bouncing ball orbits described above. The better agreement for $R = 0.3$ is qualitatively explained by the pruning of bouncing ball manifolds and the reduction of their measure as $R$ increases.

As a representative of the bilinear statistics (which are based on the two–point correlation of the spectral density) we show $\Sigma^2(l)$ [20] for $R = 0, 0.2, 0.3$ and Dirichlet boundary conditions on the planes (fig. 4). The energy interval for which $\Sigma^2(l)$ is calculated is taken to be $133 \leq k \leq 146$, such that $\bar{d}(E)$ does not change significantly within this interval. For the chaotic cases, $R = 0.2, 0.3$, the departures from the universal GOE curve occur earlier than predicted by Berry [21]. Also, in contrast with generic chaotic systems, $\Sigma^2(l)$ oscillates around a saturation value which is larger than the GOE value. These deviations are more pronounced for $R = 0.2$ than for $R = 0.3$, which is a clear indication of the effect of the non–generic bouncing ball manifolds, that are larger both in number and in measure in the $R = 0.2$ case. The bouncing ball manifolds attain their full strength in the $R = 0$ (integrable) case, for which we observe an initial slope which is much steeper than the value 1 predicted for Poisson statistics. This is due to number–theoretic (NT) degeneracy of the energy levels, which depends on the dimensionality. Taking this degeneracy into account, the initial slope of $\Sigma^2(l)$ for $R = 0$ was calculated to be $0.0323\,kL$, which agrees well with the quantum data. (This estimate is based upon a numerical evaluation of the degeneracy, and was checked up to $kL = O(10^3)$.) The expression of the leading term in the $k$–expansion of the saturation value ($R = 0$) is semiclassically estimated by $1.36\,10^{-4}\,(kL)^2(\ln(kL) - 0.96)$, which is due to both the large contributions of the bouncing balls in three dimensions, and a number theoretical degeneracy of their lengths. This saturation value was found to agree with the quantum results for very large $k$. For the case shown in fig. 4, $k$ is not large enough, and the asymptotic value is not reached.

To summarize, the quantization of the 3D Sinai billiard using the KKR method became feasible due to the high symmetry of the billiard, and the resulting block–diagonal structure of the Hamiltonian. This high symmetry results in non–generic bouncing ball periodic manifolds. The infinite number of the bouncing ball manifolds were shown to dominate the length spectrum in a way that have important qualitative differences from the 2D case. Genuine 3D geometrical effects give rise to dramatic reduction of the semiclassical contribution of the shortest unstable periodic orbit. The quantum short range level statistics were shown to agree quite well with GOE results, in accordance with the common experience with systems that have time–reversal symmetry. However, important deviations have been observed in the $\Sigma^2$ statistics which were attributed to the bouncing ball manifolds.

## ACKNOWLEDGMENTS

We are grateful to M. Sieber, E. Bogomolny, M.V. Berry and H. Schanz for their interest and for helpful discussions. The research was supported by the Binational U.S.–Israel Science Foundation (BSF) and the Minerva Center for Non–Linear Studies of Complex Systems.

FIGURES

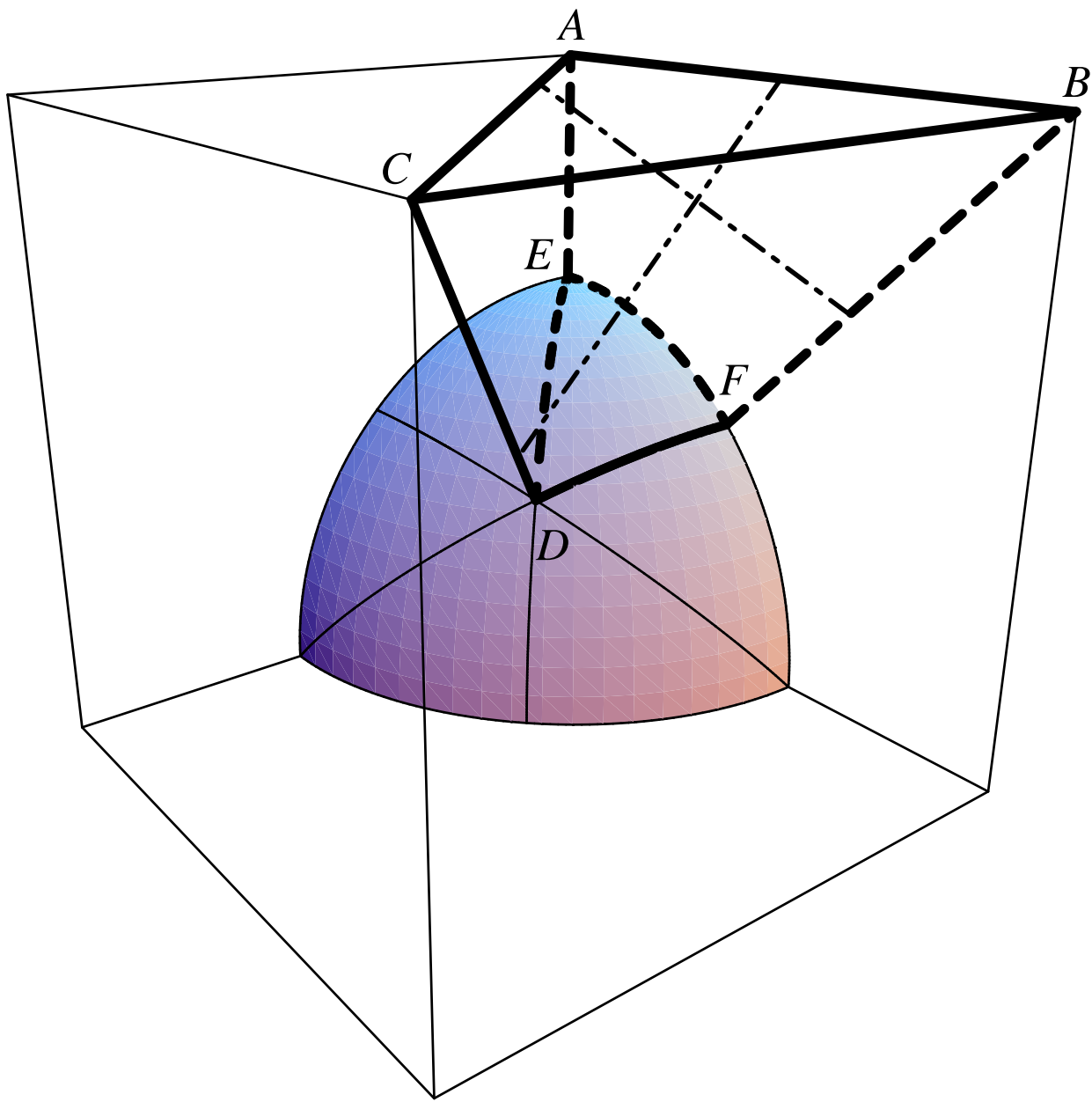

FIG. 1. Desymmetrized 3D Sinai billiard. The billiard is indicated by boldface edges, and its corners are marked by letters. Chained line: shortest isolated periodic orbit of length $L/\sqrt{3}$. Double chained: isolated periodic orbit of length $L/\sqrt{2}$.

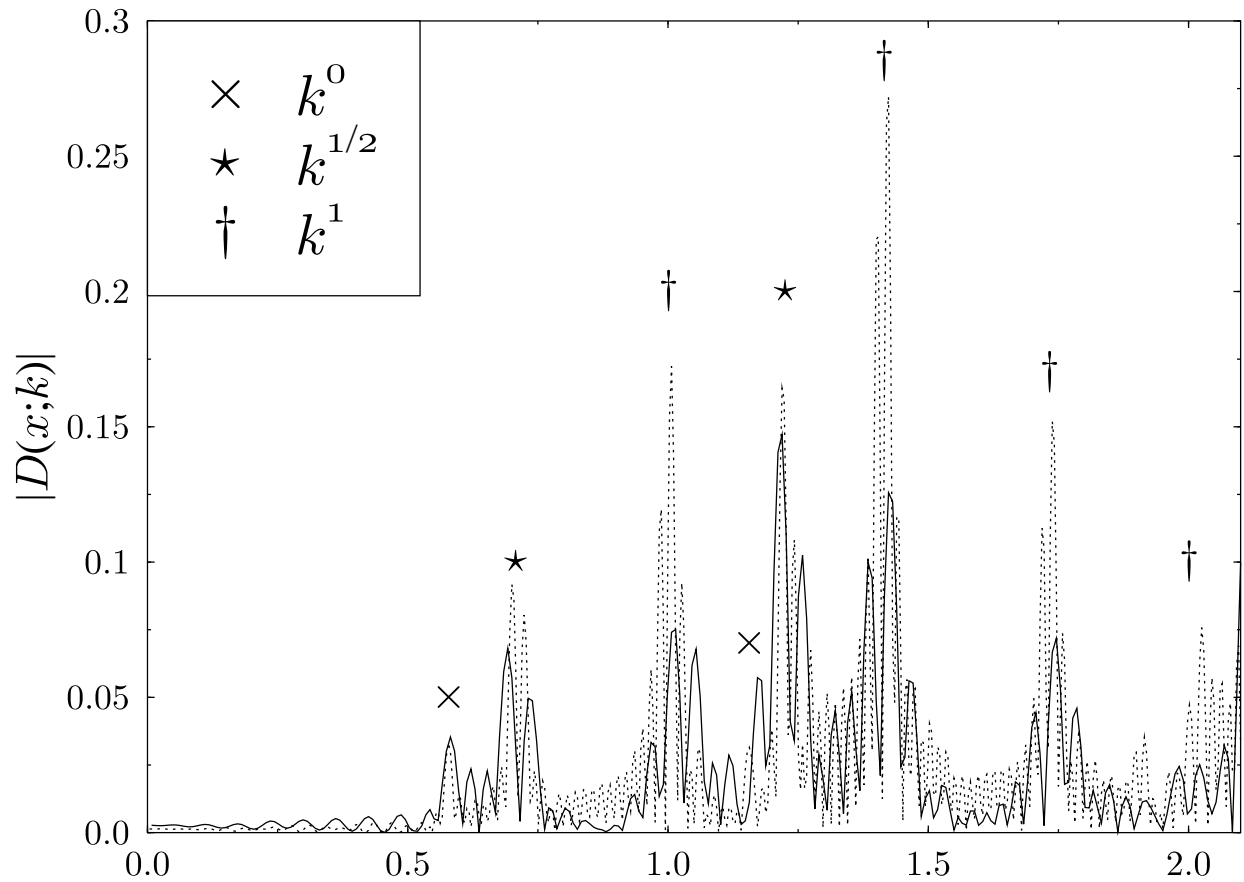

FIG. 2. Absolute value of length spectrum for the $R = 0.2$ (DD) spectrum. Full line: $k = 100$, dashed line: $k = 200$. The first 8 bouncing ball orbits are indicated together with their relative contribution to $D(x;k)$.

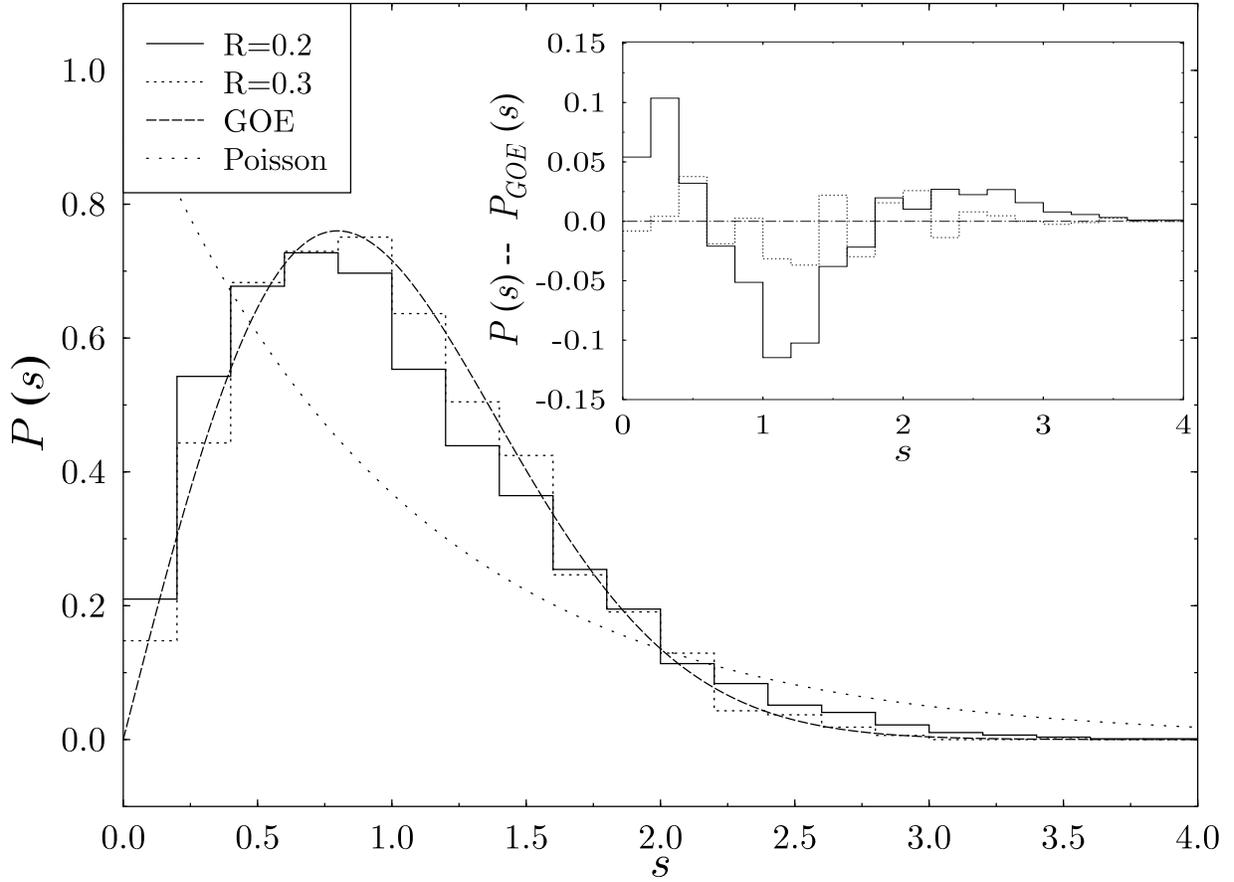

FIG. 3. Nearest neighbor distribution $P(s)$. For each $R$, data are averaged over available spectra with different boundary conditions. Inset: difference between $P(s)$ and $P_{GOE}(s)$.

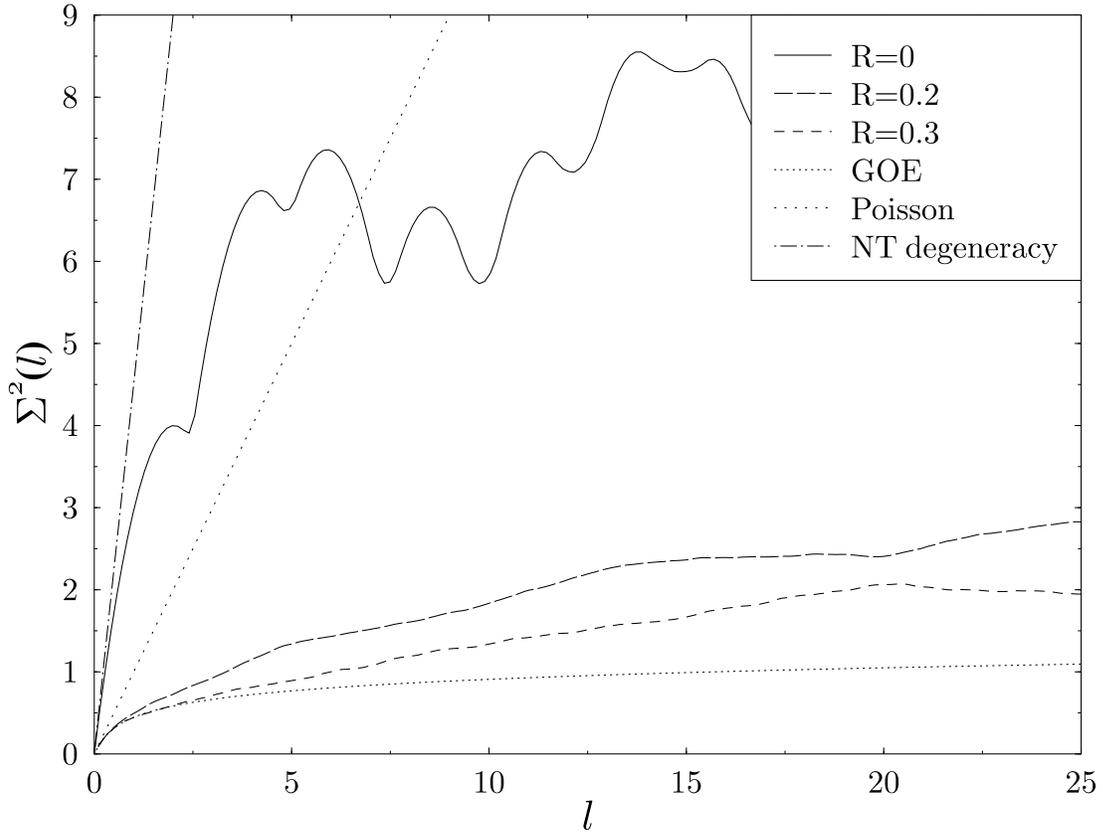

FIG. 4. $\Sigma^2(l)$ bilinear statistics for $133 \leq k \leq 146$. For $R = 0.2, 0.3$ we averaged over two spectra corresponding to Neumann and Dirichlet boundary conditions on the sphere.